\documentstyle[prl,aps,epsfig]{revtex}         
\begin{document}
\draft               
\twocolumn[\hsize\textwidth\columnwidth\hsize\csname @twocolumnfalse\endcsname

\title{Spatial solitons in an optically pumped semiconductor microresonator}
\author{V. B. Taranenko, C. O. Weiss}
\address{Physikalisch-Technische Bundesanstalt 38116 Braunschweig, Germany}
\maketitle
\begin{abstract}
We show experimentally and numerically the existence of stable
spatial solitons in an optically pumped semiconductor
microresonator. We demonstrate that the pump substantially
reduces the light intensity necessary to sustain the solitons and
thereby reduces destabilizing thermal effects. We demonstrate
coherent switching on and off of bright solitons. We discuss
differences between pumped and unpumped below-bandgap-solitons.
\end{abstract}
 \vskip1pc ]
Spatial solitons in semiconductor microresonators
\cite{tag:1,tag:2} are self-trapped light beams which can form
due to both transverse (e.g., self-focusing) and longitudinal
(nonlinear resonance \cite{tag:3}) nonlinear effects. These
nonlinear effects can act in the same sense or oppositely, with
the consequence of reduced soliton stability in the later case.
Thus choice of nonlinear resonator parameters suited best for
sustaining stable solitons is important. The best-suited schemes
are unpumped microresonator excited above bandgap \cite{tag:4}
and optically pumped microresonator excited below bandgap
\cite{tag:5}.

Here we pump a semiconductor microresonator optically in a range
up to the lasing threshold and study formation of switched
structures. We show experimentally and numerically the existence
of the resonator solitons in the pumped microresonators. We
demonstrate that the pumping substantially reduces the light
intensity necessary to sustain and switch the solitons
(diminishing thermal problems as a side effect) so that
semiconductor laser diodes are sufficient for sustaining
solitons.  We discuss differences between pumped and unpumped
solitons below bandgap.\\

\begin{figure}[htbf] \epsfxsize=80mm
\centerline{\epsfbox{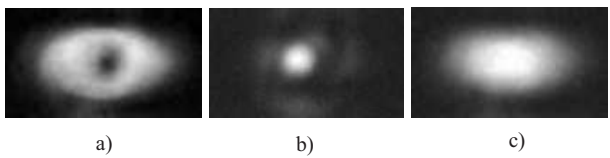}} \vspace{0.7cm} \caption{Intensity
snapshots of typical structures observed in reflection from
pumped (below transparency) semiconductor microresonator
illuminated near resonance showing bright (a) and dark (b)
soliton. \\
The illuminating beam from the laser diode has an elliptical
shape (c).}
\end{figure}

We pump a bistable quantum-well (GaAs/GaAlAs) microresonator
using either a tunable Ti:sapphire laser or a high power
multimode laser diode. The pumped area of the resonator is
illuminated additionally by a focused beam from a single-mode
laser diode that provides quite large Fresnel number ($\sim$100)
and near resonant illumination of the resonator. The main control
parameters in the experiment are the resonator detuning and
intensities of the illumination and pump.

Observations are done in reflection since the substrate of the
microresonator structure is opaque at the working wavelengths:
dark switched structures in reflection correspond to bright ones
in transmission and vice versa. Switched structures formed in the
illuminated beam cross section were monitored in the plane of the
microresonator in two ways: (i) A CCD camera with electro-optical
shutter recorded 2D snapshots of switched structures. (ii) A fast
and small aperture photo detector monitored local dynamics.\\

Switched structures observed (Fig.~1 a,b) manifest themselves as
resonator solitons: 1) they are of the size ($\sim$10 $\mu$m)
expected for such solitons \cite{tag:1}; 2) they are round spots
whose size and shape are independent on the intensity and shape
of the illuminating beam (e.g., elliptical beam shape as shown in
Fig.~1c); 3) they are robust against perturbations of the
illuminating light intensity; 4) they are bistable, i.e. they can
be switched on (Fig.~2a) and off (Fig.~2b) by sharply focused
address pulses.

\begin{figure}[htbf]
\epsfxsize=80mm \centerline{\epsfbox{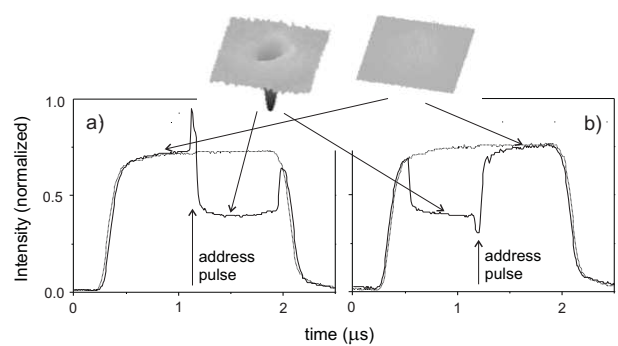}} \vspace{0.7cm}
\caption{Recording of switching-on (a) and switching-off (b) of a
bright soliton. Vertical arrows mark the application of address
pulses. Dotted traces: incident intensity. The insets show
soliton and unswitched state in 3D representation.}
\end{figure}

This soliton nature of the observed switched structures is
supported by numerical simulations of the intracavity field
structures (in 2D) using the model equations for the pumped
semiconductor microresonator driven by a plane wave \cite{tag:5}.
Fig.~3 shows typical examples of calculated resonator solitons
below bandgap. Bright solitons have a large existence range in
the pumped case (Fig.~3b), dark solitons exist, though with
smaller range of stability, in the unpumped case (Fig.~3a).
Parameter domains of existence of resonator solitons are related
to those of optical bistability for plane waves, are shown in
Fig.~3.

\begin{figure}[htbf]
\epsfxsize=85mm \centerline{\epsfbox{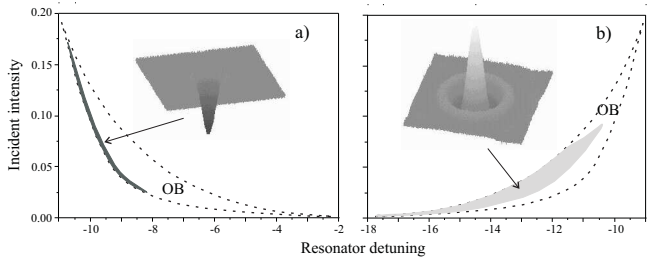}} \vspace{0.7cm}
\caption{Results of numerical simulations of below bandgap
(purely dispersive) solitons using the model equations for
unpumped (a) and pumped above "transparency" point (b)
microresonator. Insets are dark (a) and bright (b) solitons.
Shaded areas are domains of existence of resonator solitons.
Areas limited by dashed lines are optical bistability domains for
plane waves. }
\end{figure}

Analysis shows that increase of the pump intensity leads to
shrinking of the resonator solitons' existence domain and
shifting towards low intensity of the light sustaining the
solitons (Such reduction of the sustaining light intensity was
observed experimentally in \cite{tag:5}).

When the pump intensity approaches the transparency point of the
semiconductor material, the resonator solitons' domain of
existence disappears. It reappears above the transparency point.
In the experiment we have quite strong contribution of the
imaginary part (absorption/gain) of the complex nonlinearity at
the working wavelength (854 nm). Therefore the transparency point
is very close to the lasing threshold so that inversion without
lasing is difficult to realize.

\begin{figure}[htbf]
\epsfxsize=80mm \centerline{\epsfbox{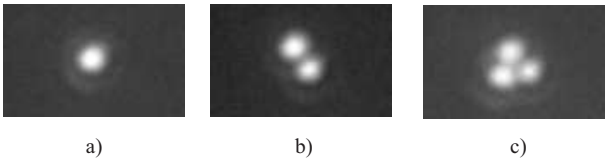}} \vspace{0.7cm}
\caption{Intensity snapshots of typical beam structures at
optical pump intensities slightly above lasing threshold (pump
increases from (a) to (c)). }
\end{figure}

Slightly above threshold we observe in presence of illumination
structures (Fig.~4) reminiscent of the solitons in electrically
pumped resonators \cite{tag:6}.\\

In summary, optically pumped semiconductor resonators are well
suited for sustaining solitons below bandgap:(i) background light
intensity necessary to sustain and switch resonator solitons is
substantially reduced by the pumping and therefore destabilizing
thermal effects are minimized, (ii) above the transparency point
only the dispersive part of semiconductor nonlinearity stabilizes
a soliton, therefore the domain of existence of "below bandgap"
(purely dispersive) bright solitons can be quite large. Moreover
optical as opposed to electrical pumping allows more homogeneous
pumping conditions \cite{tag:7}. This suggests that optically
pumped resonators lend themselves more readily for localization
and motion control of solitons then electrically pumped ones. \\

Acknowledgment\\ This work was supported by Deutsche
Forschungsgemeinschaft under grant We743/12-1.

\end{document}